\documentclass{eas}
\usepackage{graphicx}
\begin{document}

\title{Satellites in the Local Group\\ \smallskip and Other Nearby Groups} 
\author{Eva K.\ Grebel}\address{Astronomisches Rechen-Institut,
Zentrum f\"ur Astronomie der Universit\"at Heidelberg,
M\"onchhofstr.\ 12--14, 69120 Heidelberg, Germany}
%
%
\begin{abstract}
In recent years the census of known satellites in our own Local Group and
in nearby galaxy groups has increased substantially due to sensitive
wide-area surveys. In the Local Group these surveys have more than doubled
its known galaxy content and extended the galaxy luminosity function to very
faint total magnitudes. Deep ground-based imaging and spectroscopic
observations as well as high-resolution imaging with the Hubble Space
Telescope have revolutionized our understanding of the chemical evolution
and star formation histories of the satellites. We often find long-lasting
star formation episodes with low star formation efficiencies. There is
evidence for localized, stochastic enrichment, and recent searches are now
beginning to uncover even extremely metal-deficient stars. In many
satellites evidence for two or more distinct stellar subpopulations is
found. Differing fractions of old populations have been detected in all
satellites studied in sufficient detail so far. Kinematic measurements
support a picture in which satellites are dark-matter dominated, although
recent results indicate that the proposed common mass scale for dwarf 
spheroidal galaxies may not hold
for very low-mass satellites.  When considering satellite ensembles, we
find global morphology--distance and gas-content--distance relations in all
groups studied thus far, but individual star formation histories 
also strongly depend on a given satellite's intrinsic properties.
\end{abstract}
\maketitle
\section{Introduction}

As noted by Karachentsev (\cite{Kara05}), 85\% of the galaxies in the
Universe are located outside of rich galaxy clusters.  About half of these
galaxies are found in groups, and this fraction increases when also
counting galaxies in unbound ``clouds'' and filaments.  Such group
environments are thus typical for the regions in which the majority of 
galaxies currently live. They are also excellent laboratories to study
galaxy evolution throughout the wide range of commonly found environments.

In this contribution I will focus on nearby groups of galaxies within about
5 Mpc, which are close enough to permit us to resolve individual stars well
below the tip of the red giant branch with only a modest investment of
observing time with high-resolution cameras such as the ones aboard the
Hubble Space Telescope (HST).  These galaxies in the ``Local Volume'' have
been targeted by several major observing campaigns. These include our
200-orbit HST snapshot program (Grebel \cite{Grebel00a}) and the ``ACS
Nearby Galaxy Survey Treasury'' (ANGST, Dalcanton {\em et al.}
\cite{Dalcanton09}) for resolved stellar populations, as well as
ground-based broad-band and narrow-band imaging campaigns for integrated
properties (e.g., Bremnes {\em et al.} \cite{Bremnes00}; Jerjen {\em et
al.} \cite{Jerjen00}; Karachentsev {\em et al.} \cite{Karachentsev00};
Barazza {\em et al.} \cite{Barazza01}; Makarova {\em et al.}
\cite{Makarova02a}, \cite{Makarova05}; Chiboucas {\em et al.}
\cite{Chiboucas09}; C\^ot\'e {\em et al.} \cite{Cote09}; Fingerhut {\em et
al.} \cite{Fingerhut10}), ground-based studies of H\,{\sc i} content and
kinematics (e.g., Banks {\em et al.} \cite{Banks99}; C\^ot\'e {\em et al.}
\cite{Cote97}, \cite{Cote00}; Bouchard {\em et al.} \cite{Bouchard05}), and
of (nebular) abundances (e.g., Richer {\em et al.} \cite{Richer98};
Skillman {\em et al.} \cite{Skillman03}; Lee {\em et al.} \cite{Lee03},
\cite{Lee07}; Kniazev {\em et al.} \cite{Kniazev05}, \cite{Kniazev08};
Croxall {\em et al.} \cite{Croxall09}).  While these programs provide an
increasingly elaborate picture of dwarf galaxy properties and evolution in
nearby groups, the most detailed information is available for dwarf
galaxies in the Local Group, particularly Milky Way companions, where we
can even resolve stars below the oldest main-sequence turn-offs.  Thus,
special emphasis will be placed on the Local Group in this review.

Our focus on the Local Group and other nearby groups within 5 Mpc implies
that certain types of groups -- in particular rich, fossil, or compact
groups (see, e.g., Lee {\em et al.} \cite{Lee04}) -- are excluded because
we lack nearby examples.  Nonetheless, we have a range of different group
environments nearby, including strongly interacting and more quiescent
groups, poor groups, and very loose groups or ``clouds'' (see also Tully
\cite{Tully87}; Grebel \cite{Grebel07}).  The most commonly occurring dwarf
galaxies in these nearby groups are dwarf elliptical (dE) and dwarf
spheroidal (dSph) galaxies among the early-type dwarfs and dwarf irregular
(dIrr) galaxies among the late-type dwarfs.  Ultra-compact (e.g.,
Evstigneeva {\em et al.} \cite{Evstigneeva07}; Rejkuba {\em et al.}
\cite{Rejkuba07}) and compact dwarfs (e.g., Ziegler \& Bender
\cite{Ziegler98}) and blue compact dwarfs (Gil de Paz {\em et al.}
\cite{GildePaz03}; Kniazev {\em et al.} \cite{Kniazev03}; Richer {\em et
al.} \cite{Richer01}) are rare.  DEs and dSphs are usually satellites of
the dominant, massive galaxies in groups, while dIrrs tend to be found at
larger distances and may be bound to the group as a whole rather than to an
individual high-mass galaxy.  For a review of the properties of different
types of dwarf galaxies, see Grebel (\cite{Grebel01}).  -- The interacting
M81 group may even contain genuine tidal dwarf galaxies (Makarova {\em et
al.} \cite{Makarova02b}; Croxall {\em et al.} {\cite{Croxall09}; see
Bournaud \cite{Bournaud10} for a general review).

\section{Morphological Segregation and Environment}

Nearby galaxy groups within 5 Mpc range from evolved groups like the
Centaurus A group (e.g., Karachentsev {\em et al.} \cite{Kara02a},
\cite{Kara07}; C\^ot\'e {\em et al.} \cite{Cote09}), the IC\,342/Maffei
group (e.g., Buta \& McCall \cite{Buta99}; Karachentsev {\em et al.}
\cite{Kara03a}; Fingerhut {\em et al.} \cite{Fingerhut07}) and the
interacting M81 group (e.g., Yun {\em et al.} \cite{Yun94}; Karachentsev
{\em et al.} \cite{Kara02b}; Chynoweth {\em et al.} \cite{Chynoweth08};
Chiboucas {\em et al.} \cite{Chiboucas09}) to little evolved, extended
filaments or ``clouds'' such as the Sculptor and the Canes Venatici I group
(e.g., Karachentsev {\em et al.} \cite{Kara03b}, \cite{Kara03c}).   In
addition, several sparse, possibly bound `dwarf groups'' have been
identified in the Local Volume (e.g., van den Bergh \cite{vdBergh99a}; Tully
{\em et al.} \cite{Tully06}), whose most massive galaxy is usually an
irregular.

The clouds or filaments contain primarily gas-rich, late-type galaxies that
do not yet seem to have experienced many interactions or significant
stripping of their gaseous material.  The more evolved prominent galaxy
groups typically contain two or more massive galaxies, each of which is
surrounded by an entourage of lower-mass satellites.  Similar 
as in galaxy clusters (e.g., Oemler \cite{Oemler74}; Dressler
\cite{Dressler80}; Lisker {\em et al.} \cite{Lisker07}), where the fraction
of early-type galaxies increases with galaxy density, in these groups as
well as in the Local Group a clear morphology-density relation is observed
(e.g., Einasto et al.\ \cite{Einasto74}; van den Bergh
\cite{vdBergh99b}; Grebel \cite{Grebel99}, \cite{Grebel00b},
\cite{Grebel05}; and the above references).  In particular, a pronounced
increase in the number of early-type dwarfs is found within $\sim 300$ kpc
around the massive galaxies in the more evolved groups.  Environmental
effects such as ram pressure, tidal forces, and galaxy harassment are
believed to be responsible for the morphological segregation among dwarf
galaxies (e.g., van den Bergh \cite{vdBergh94}; Grebel {\em et al.}
\cite{Grebel03}; Kravtsov {\em et al.} \cite{Kravtsov04}; Mayer {\em et
al.} \cite{Mayer06}; D'Onghia {\em et al.} \cite{DOnghia09}; Grcevich \&
Putman \cite{Grcevich09}).  Also in more distant groups there is clear
evidence both for morphological and for luminosity segregation (e.g., 
Girardi {\em et al.} \cite{Girardi03}).  

The morphology-density relation can be expressed as a morphology-radius
relation or an H\,{\sc i} mass-radius relation (e.g., Einasto {\em et al.}
\cite{Einasto74}; Grebel {\em et al.} \cite{Grebel03}, Fig.\ 3).  The upper
limits of the H\,{\sc i} masses for dwarf galaxies within $\sim 270$ kpc
from the Milky Way or M31 are typically less than $10^4$~M$_{\odot}$
(Grcevich \& Putman \cite{Grcevich09}), and only the more massive dwarfs
manage to retain some of their neutral gas at these close distances.  Thus
far there is no evidence for significant amounts of ionized gas in low-mass
dSphs (e.g., Gallagher {\em et al.} \cite{Gallagher03}).  These findings
contrast with the low-density, extended Sculptor group filament, where many
of the early-type dwarfs were found to have H\,{\sc i} masses of several
$10^5$~M$_{\odot}$ (Bouchard {\em et al.} \cite{Bouchard05}), seemingly
having been able to retain gas from normal stellar outflows in the absence
of external stripping.  On the other hand, ram pressure stripping appears
to be at work in the higher-density Cen A group (Bouchard {\em et al.}
\cite{Bouchard07}). 

\section{Faint Satellite Luminosities, Masses, and the Substructure Problem}

One of the long-standing unsolved questions in cosmology is the
substructure problem -- the problem that cosmological simulations predict
up to two orders of magnitude more small dark matter halos than the number
of actually observed satellite galaxies (e.g., Klypin {\em et al.}
\cite{Klypin99}; Moore {\em et al.} \cite{Moore99}; Kravtsov
\cite{Kravtsov10}).  Many scenarios have been suggested in order to resolve
this apparent discrepancy. Among others, these scenarios include the
existence of dark matter halos without baryons due to early
photoevaporation, possibly during reionization or because of various
feedback effects (e.g., Bullock {\em et al.} \cite{Bullock00}; Madau {\em
et al.} \cite{Madau08}; Wadepuhl \& Springel \cite{Wadepuhl11}), a
substantial underestimate of the observed circular velocities of the
satellites (e.g., Stoehr {\em et al.} \cite{Stoehr02}; Pe\~narrubia {\em et
al.} \cite{Penarrubia08}), or an observational bias in detecting faint
satellites (e.g., Tollerud {\em et al.} \cite{Tollerud08}; Willman
\cite{Willman10}).  

Indeed the galaxy census of the Local Group and of nearby groups is
significantly incomplete toward the faint end of the galaxy luminosity
function.  In recent years, a considerable number of new satellites was
discovered both in the Local Group (e.g., Zucker {\em et al.}
\cite{Zucker04a}, \cite{Zucker06a}, \cite{Zucker06b}, \cite{Zucker07};
Belokurov {\em et al.} \cite{Belokurov06}, \cite{Belokurov07},
\cite{Belokurov10}; Willman {\em et al.} \cite{Willman05}; Martin {\em et
al.} \cite{Martin06}, \cite{Martin09}; Chapman {\em et al.}
\cite{Chapman07}; Walsh {\em et al.} \cite{Walsh07}; Majewski {\em et al.}
\cite{Majewski07}; McConnachie {\em et al.} \cite{McConnachie08}) and in
nearby groups (e.g., C\^ot\'e {\em et al.} \cite{Cote97}; Karachentseva \&
Karachentsev \cite{Karachentseva98}; Banks {\em et al.} \cite{Banks99};
Chiboucas {\em et al.} \cite{Chiboucas09}).  More sensitive surveys and
wider sky coverage should reveal even more faint dwarfs (see Tollerud {\em
et al.} \cite{Tollerud08}), alleviating the ``missing satellite problem''.
On the other hand,  the higher resolution of modern cosmological
simulations tends to predict even more small dark matter halos, worsening
the discrepancy again (Madau {\em et al.} \cite{Madau08}; Bovill \& Ricotti
\cite{Bovill09}).  

Most of the newly discovered dwarfs are dim dSphs. These include the
``ultra-faint'' dSphs (mostly Milky Way satellites), the faintest galaxies
known with luminosities down to $\sim 10^3$~L$_{\odot}$.  The new Local
Group dwarfs are mainly dominated by old populations and are typically
quite metal-poor (e.g., Whiting {\em et al.} \cite{Whiting99}; Grebel \&
Guhathakurta \cite{GG99}; Harbeck {\em et al.} \cite{Harbeck04},
\cite{Harbeck05}; Simon \& Geha \cite{Simon07}; Kirby {\em et al.}
\cite{Kirby08}; Koch {\em et al.} \cite{Koch08a}; Kalirai {\em et al.}
\cite{Kalirai09}; Frebel {\em et al.} \cite{Frebel10a}; Ad\'en {\em et al.}
\cite{Aden11}, and references above).  Such dwarf galaxies that formed most
of their stars at $z\sim 8$ could represent ``fossils'' of the era of
reionization (e.g., Gnedin \& Kravtsov \cite{Gnedin06}; Bovill \& Ricotti
\cite{Bovill09}).  While the current detections in other groups or even
around M31 do not yet reach as faint a luminosity as in the new Milky Way
companions, the faint-end slope of the cumulative luminosity functions in
nearby groups appears to be rather similar, typically of the order of $-1.3
\pm 0.1$, apparently independent of environment (Chiboucas {\em et al.}
\cite{Chiboucas09}).  

Interestingly, the radial velocity dispersion profiles of dSphs are fairly
flat as a function of radius and reveal high velocity dispersions even at
large radii, which is usually interpreted as an indication of a high dark
matter content (e.g., Wilkinson {\em et al.} \cite{Wilkinson04}; Koch {\em
et al.} \cite{Koch07a}, \cite{Koch07b}; Gilmore {\em et al.}
\cite{Gilmore07a}; Walker {\em et al.} \cite{Walker07}; Ural {\em et al.}
\cite{Ural10}).  In fact, it has been suggested that all dSph galaxies may
share a common halo mass (e.g., Gilmore {\em et al.} \cite{Gilmore07a};
Strigari {\em et al.} \cite{Strigari08}; Walker {\em et al.}
\cite{Walker09}; Wolf {\em et al.} \cite{Wolf10}), and that perhaps even
the central surface density of dark matter halos is approximately constant
and independent of galaxy luminosity (e.g., Donato {\em et al.}
\cite{Donato09}).  However, there are also indications of deviations from a
uniform enclosed mass out to a constant radius:  Some measurements suggest
that there is a decline in halo mass with decreasing luminosity after all.
Careful removal of interloper stars in sparse stellar systems leads to
lower velocity dispersions and masses (Ad\'en {\em et al.} \cite{Aden09a},
\cite{Aden09b}).  Also, there may be intrinsic differences between M31 and
Milky Way dSph satellites, and some some of the ultra-faint dSphs may be
experiencing tidal disruption (e.g., Kalirai {\em et al.} \cite{Kalirai10};
Collins {\em et al.} \cite{Collins10}; Simon \& Geha \cite{Simon07}, but
see Pe\~narrubia {\em et al.} \cite{Penarrubia08} for a different view).  

Generally, caution is advisable when interpreting morphological features as
signatures of tidal disruption (Mu\~noz {\em et al.} \cite{Munoz08}).  Thus
far no evidence of Galactic dSphs being unbound tidal remnants with a large
line-of-sight depth has been found (Klessen {\em et al.} \cite{Klessen03}).
In any case, the existing stellar radial velocity profiles probably still
only trace the dwarfs' inner regions; at larger distances a decline is
expected (e.g., Prada {\em et al.} \cite{Prada03}).  Deeper surveys do
indeed occasionally reveal that even the luminous extent of dwarf galaxies
is larger than previously assumed (e.g., Odenkirchen {\em et al.}
\cite{Odenk01}; Komiyama {\em et al.} \cite{Komiyama03}; Kniazev {\em et
al.} \cite{Kniazev09}).    

Velocity dispersion profiles appear to indicate a preference for cored
profiles in dSphs (e.g., Gilmore {\em et al.} \cite{Gilmore07b}; Battaglia
{\em et al.} \cite{Battaglia08}), although cuspy dark matter profiles are
not excluded.  Observations also seem to support constant dark matter
densities in the central regions of galaxies (e.g., Donato {\em et al.}
\cite{Donato09}), whereas cosmological simulations predict steeply rising
profiles.  This ``core/cusp'' problem is another unsolved problem in modern
cosmology (see de Blok \cite{deBlok10} for a review).  A promising solution
of this long-standing problem may be that initially cuspy dark matter
profiles become cored simply as a consequence of star formation and
feedback, resolving the contradiction (e.g., Pasetto {\em et al.}
\cite{Pasetto10}).

\section{Star Formation Histories and Chemical Evolution}

\subsection{Global star formation histories}

Global star-formation rates of galaxies in the present-day Universe tend to
increase with decreasing stellar mass, but show increasing scatter when
entering the low-luminosity dwarf regime (Bothwell {\em et al.}
\cite{Bothwell09}).  Many late-type dwarf galaxies have a sufficiently high
H\,{\sc i} content to continue to form stars for another Hubble time, but
still show only very low star-formation rates, which may be due to very low
gas densities that increase the H\,{\sc i} consumption time scales (see
also Hunter \cite{Hunter97}).  Low-efficiency star formation may, however,
evade detection as it would be difficult to trace with commonly used
techniques such as H$\alpha$ emissivity (Bothwell {\em et al.}
\cite{Bothwell09}).  Apart from the range of present-day star formation
properties in late-type dwarfs, the scatter in star-formation rates at the
low-mass end of the baryonic mass distribution is further increased by the
presence of quiescent, gas-deficient early-type dwarfs.   

In the Local Group no two dwarf galaxies share the same star formation
history, not even within the same morphological type (Grebel
\cite{Grebel97}, \cite{Grebel99}; see also Orban {\em et al.}
\cite{Orban08}).  Despite their differences, dwarfs share many common
evolutionary trends and follow similar scaling relations (e.g., Sharina
{\em et al.} \cite{Sharina08}).  Apart from morphological segregation, the
star formation rates and star formation histories of dwarfs in the Local
Group and in nearby groups do not show a clear correlation with distance
from the closest massive primary (e.g., Weisz {\em et al.} \cite{Weisz08};
C\^ot\'e {\em et al.} \cite{Cote09}; Crnojevi\'c {\em et al.}
\cite{Crnojevic10}; Lianou {\em et al.} \cite{Lianou10}).  This suggests
that intrinsic properties rather than environment may govern the evolution
of dwarf galaxies (see also Revaz {\em et al.} \cite{Revaz09}).  However,
the orbits of these dwarfs are unknown.  Models considering realistic
orbits ({\em for Milky Way companions}) support the triggering of star 
formation activity in dwarfs through encounters (e.g., Pasetto {\em et al.}
\cite{Pasetto11}).  

\subsection{Old stellar populations}

All Local Group dwarfs studied in detail thus far show evidence for very
old populations (e.g., Grebel \& Gallagher \cite{GG04}) as demonstrated by
the presence of old, age-dateable main-sequence turn-offs in field
populations or globular clusters, horizontal branches, and/or RR Lyrae
stars.  The fraction of old stars varies considerably from galaxy to
galaxy.  An example of a dIrr galaxy with a very modest old population is
the Small Magellanic Cloud (SMC):  Its only globular cluster is about two
billion years younger than typical old globulars in the Large Magellanic
Cloud, Fornax, Sagittarius, and the Milky Way (e.g., Glatt {\em et al.}
\cite{Glatt08a}).  RR Lyrae stars were detected in the SMC in much smaller
numbers than in the neighboring Large Magellanic Cloud (Soszy\'nski {\em
et al.} \cite{Soszynski10}).  In a few other Local Group
dIrrs the existence of an ancient population was questioned until finally
RR Lyrae stars were detected (e.g., Dolphin {\em et al.} 
\cite{Dolphin02}).  In contrast, early-type dwarfs typically contain 
rather prominent old populations. 

{From} their analysis of 60 Local Volume dwarfs with color-magnitude
diagrams reaching the upper red giant branch, Weisz {\em et al.}
(\cite{Weisz11}) infer that, on average, dwarf galaxies formed more than
50\% of their stars by $z\sim2$ (see also Crnojevi\'c {\em et al.}
\cite{Crnojevic11}).  {\em Unambiguous}\ evidence for ancient
populations beyond the Local Group, however, so far only exists for two
dwarfs in the Sculptor group, where blue horizontal branches and RR Lyrae
stars were detected (Da Costa {\em et al.} \cite{DaCosta10}).  

\subsection{Abundance spreads and inhomogeneous enrichment}

Dwarf galaxies typically reveal indications of extended episodes of star
formation leading to considerable abundance spreads (of one dex in [Fe/H]
or more) even in purely old populations (see, e.g., Shetrone {\em et al.}
\cite{Shetrone01}; Grebel {\em et al.} \cite{Grebel03}; Cohen \& Huang
\cite{Cohen10}).  While we lack stellar spectroscopy for galaxies beyond the
Local Group, the broad red giant branches of dwarfs in nearby groups
suggest the existence of considerable metallicity spreads there as well
(e.g., Sharina {\em et al.} \cite{Sharina08}; Crnojevi\'c {\em et al.}
\cite{Crnojevic10}; Lianou {\em et al.} \cite{Lianou10}) and can be
translated into photometric metallicity distribution functions.  For Local
Group satellites a growing number of spectroscopic metallicity distribution
functions is available (e.g., Helmi {\em et al.} \cite{Helmi06}; Koch {\em
et al.} \cite{Koch06}, \cite{Koch07b}, \cite{Koch07c}; Kirby {\em et al.}
\cite{Kirby08}), confirming large spreads.

In some nearby galaxies evidence for metallicity spreads has been found
even within coeval populations (e.g., Venn {\em et al.} \cite{Venn03};
Kniazev {\em et al.} \cite{Kniazev05}; Koch {\em et al.} \cite{Koch07c}).
This is particularly pronounced in the SMC, where metallicity spreads seem
to exist at any given age (Glatt {\em et al.} \cite{Glatt08b}).  There is
also evidence for inhomogeneous enrichment among stars of the same
metallicity.  Individual element abundance ratios may permit one to even
trace individual supernova events (e.g., Sadakane {\em et al.}
\cite{Sadakane04}; Koch {\em et al.} \cite{Koch08a}, \cite{Koch08b}; Ad\'en
{\em et al.} \cite{Aden11}).  

Abundance inhomogeneities are usually attributed to slow, stochastic,
localized star formation with low star formation efficiency (e.g.,
Marcolini {\em et al.} \cite{Marcolini08}).  These models also predict the
usually observed asymmetric metallicity distribution functions with an
extend low-metallicity tail, a slow rise toward higher metallicities and
subsequently a rapid fall-off.  Alternatively, there are successful models
in which star formation rates and galactic wind efficiencies are chosen to
reproduce the observed metallicity distributions (e.g., Lanfranchi \&
Matteucci \cite{Lanfranchi07}, \cite{Lanfranchi10}).  

Overall, the dwarf galaxies in the Local Group as well as those in nearby
groups follow similar metallicity-luminosity relations, and there appears
to be a general trend for dSphs being ``too metal-rich'' for their
luminosity as compared to late-type dwarfs (e.g., Richer {\em et al.}
\cite{Richer98}; Grebel {\em et al.} \cite{Grebel03}; Sharina {\em et al.}
\cite{Sharina08}).  This interesting trend may indicate that early-type
dwarfs experienced more efficient and rapid enrichment early on, while the
chemical evolution of late-type dwarfs, which essentially experienced
continuous star formation (Cignoni \& Tosi \cite{Cignoni10}), was slower
and less efficient.  As discussed in Grebel {\em et al.} (\cite{Grebel03}),
this difference makes it more difficult to turn dIrrs into dSphs. But 
low-mass transition-type galaxies may be plausible progenitors of dSphs in
terms of their stellar population properties, metallicities, and
star-formation histories.

\subsection{Stellar population gradients}

Many, but not all, dwarf galaxies exhibit population gradients.  When
gradients are present, the more metal-rich and/or younger populations are
usually found to be more centrally concentrated, while older and/or more
metal-poor populations have a more extended, smoother distribution and may
show distinct kinematics with a higher velocity dispersion (e.g., Grebel
\cite{Grebel97}; Hurley-Keller {\em et al.} \cite{Hurley-Keller99};
Zaritsky {\em et al.} \cite{Zaritsky00}; Harbeck {\em et al.}
\cite{Harbeck01}; Tolstoy {\em et al.} \cite{Tolstoy04}; Koch {\em et al.}
\cite{Koch06}; Weisz {\em et al.} \cite{Weisz08}; Leaman {\em et al.}
\cite{Leaman09}; Crnojevi\'c {\em et al.} \cite{Crnojevic10}; Lianou {\em
et al.} \cite{Lianou10}; Battaglia {\em et al.} \cite{Battaglia11}).   As
these studies show, population gradients are found both in early- and
late-type dwarfs in the Local Group and nearby groups and are
well-reproduced by models (e.g., Marcolini {\em et al.}
\cite{Marcolini08}).

\subsection{Detailed element abundance ratios and satellite accretion}

Satellite galaxies may ultimately get accreted by the massive galaxies that
they orbit.  There is mounting evidence of such events in the Local Group,
both in the form of disruption and accretion streams (e.g., Ibata {\em et
al.} \cite{Ibata94}; Newberg {\em et al.} \cite{Newberg03}; Yanny {\em et
al.} \cite{Yanny03}; McConnachie {\em et al.} \cite{McConnachie09};
Williams {\em et al.} \cite{Williams11}) and as substructure in the
Galactic or M31 halo (e.g., Bell {\em et al.} \cite{Bell08}; Zucker {\em et
al.} \cite{Zucker04b}).  Interactions are clearly also playing a role in
the M81 and the Cen A groups.  However, it remains unclear to what extent
observable satellites may have contributed to the build-up of massive
galaxies, particularly their halos (see, e.g., Unavane {\em et al.}
\cite{Unavane96}; Bullock {\em et al.} \cite{Bullock01}; Robertson {\em et
al.} \cite{Robertson05}; Font {\em et al.} \cite{Font08}).   

DSphs tend to reach solar [$\alpha$/Fe] ratios already at much lower
metallicities than Galactic halo stars (the details vary from dwarf to dwarf
depending on, e.g., the initial gas mass; Pasetto {\em et al.}
\cite{Pasetto10}).  The lower [$\alpha$/Fe] ratio at a given [Fe/H] has been
attributed to several possible factors including low star formation rates (and
hence little $\alpha$ enrichment from massive supernovae of type II), or to
efficient loss of metals and supernova ejecta due to the shallow potential
wells of the dwarfs, or/and to a larger contribution of Fe from supernovae of
type Ia (Shetrone {\em et al.} \cite{Shetrone01}).  Ultimately, the
[$\alpha$/Fe] ratio may be viewed as a tracer of the accretion time.  At low
metallicities (or early accretion) dSphs and Galactic halo stars are in very
good agreement (see also below).

Nonetheless, the difference in [$\alpha$/Fe] ratios at a given [Fe/H] in
dSphs and in the Galactic halo as well as the apparent lack of extremely
metal-poor stars in dSphs were used to argue against a significant
contribution of low-mass satellites to the build-up of the Galactic halo,
suggesting a different chemical evolution history (e.g., Shetrone {\em et
al.} \cite{Shetrone01}; Helmi {\em et al.} \cite{Helmi06}; Koch {\em et
al.} \cite{Koch06}, \cite{Koch07c}).  This picture is now changing.

Many recent studies (e.g., Koch {\em et al.} \cite{Koch08b}; Kirby {\em et
al.} \cite{Kirby08}; Aoki {\em et al.} \cite{Aoki09}; Cohen \& Huang
\cite{Cohen09}, \cite{Cohen10}; Frebel {\em et al.} \cite{Frebel10a},
\cite{Frebel10b}; Tafelmeyer {\em et al.} \cite{Tafelmeyer10}) successfully
detected very metal-deficient stars in classical and ultra-faint dSphs.
Furthermore, these studies emphasize the similarity to very metal-poor
stars in the Galactic halo, proposing that the chemical properties of
low-mass dwarf satellites are well consistent with the stellar properties
of especially the outer halo.  Moreover, Nissen \& Schuster
(\cite{Nissen10}) detected two kinematically distinct populations of halo
stars in the solar neighborhood, one with high [$\alpha$/Fe] ratios and one
with low [$\alpha$/Fe] ratios, which they attribute to objects formed in
the Milky Way and to objects accreted from dwarf galaxies.  

Forthcoming new photometric and spectroscopic surveys (including the Gaia
satellite) should help to elucidate the contribution of satellites to more
massive galaxies.  Clearly, dwarf satellites play a number of important
roles:  They are interesting in their own right, revealing a wide range of
properties that still provide a puzzle for a proper understanding of galaxy
evolution, e.g., regarding the question of ``nature'' vs.\ ``nurture''.  
They provide valuable insights into stellar evolution at low
metallicities.  They may be key objects for understanding dark matter.
They may play an important role as building blocks of more massive
galaxies.  And they are important test objects for cosmological theories,
currently still challenging our ability to understand structure formation
on small scales.



\end{document}